\documentclass[12pt]{article}
\usepackage{amsmath, amssymb}

\usepackage{amsthm}

\usepackage{amsmath,amsthm}

\usepackage{epstopdf}

\usepackage{ifthen}

\ifthenelse{\isundefined{\GeneratePdf}}
{
\usepackage[dvips]{graphicx}
}
{
\usepackage[pdftex]{graphicx}
\usepackage[pdftex]{hyperref}
}

\usepackage{framed}
\ifthenelse{\isundefined{\NoGenerateLetterSize}}
{

\ifthenelse{\isundefined{\GeneratePdf}}
{}
{
\pdfpageheight\paperheight
\pdfpagewidth\paperwidth
}
}
{}

\long\def\LongVersion#1\LongVersionEnd{}
\long\def\ShortVersion#1\ShortVersionEnd{#1}

\newcommand{\Comment}[1]{}

\LongVersion 
\LongVersionEnd 

\ShortVersion 
\ShortVersionEnd 

\ShortVersion 
\ShortVersionEnd 

\newtheorem{theorem}{Theorem}[section]
\newtheorem{lemma}[theorem]{Lemma}
\newtheorem{observation}[theorem]{Observation}
\newtheorem{corollary}[theorem]{Corollary}
\newtheorem{proposition}[theorem]{Proposition}
\newtheorem{claim}[theorem]{Claim}

\theoremstyle{definition}
\newtheorem{property}[theorem]{Property}
\newtheorem{definition}[theorem]{Definition}
\theoremstyle{plain}

\newtheorem{fact}[theorem]{Fact}
\newtheorem{example}[theorem]{Example}

\theoremstyle{definition}

\theoremstyle{plain}

\Comment{
\newtheorem{theorem}{Theorem}[section]
\newtheorem{lemma}{Lemma}[section]
\newtheorem{corollary}{Corollary}[section]
\newtheorem{claim}{Claim}[section]

\newtheorem{definition}{Definition}[section]

\newtheorem{example}{Example}[section]

\theoremstyle{definition}

\newtheorem{definition}[theorem]{Definition}
\theoremstyle{plain}
}

\DeclareMathOperator*{\argmax}{argmax}

\newcommand{\Probability}[0]{\ensuremath{\hbox{\rm I\kern-2pt P}}}

\ShortVersion 
\renewcommand{\paragraph}[1]{\par\noindent\textbf{#1}}
\ShortVersionEnd 

\newcommand{\ignore}[1]{}

\begin{document}

\title{ Economic Recommendation Systems}
\author{
Gal Bahar\thanks{
 Technion--Israel Institute of Technology}
 \and Rann Smorodisky\thanks{
 Technion--Israel Institute of Technology, rann@ie.technion.ac.il. Smorodinsky gratefully acknowledges the support of ISF grant 2016301, the joint Microsoft-Technion e-Commerce Lab, Technion VPR grants and the Bernard M. Gordon Center for Systems Engineering at the Technion.}
 \and Moshe Tennenholtz\thanks{
 Technion--Israel Institute of Technology, moshet@ie.technion.ac.il. Tennenholtz gratefully acknowledges the support of the joint Microsoft-Technion e-Commerce Lab.}}

\date{\today}

\maketitle \thispagestyle{empty}

\begin{abstract}

In the on-line Explore \& Exploit literature, central to Machine Learning, a central planner is faced with a set of alternatives, each yielding some unknown reward. The planner's goal is to learn the optimal alternative as soon as possible, via experimentation. A typical assumption in this model is that the planner has full control over the experiment design and implementation. When experiments are implemented by a society of self-motivated agents the planner can only recommend experimentation but has no power to enforce it.
Kremer et. al. \cite{Kremer2014} introduce the first study of explore and exploit schemes that account for agents' incentives. In their model it is implicitly assumed that agents do not see nor communicate with each other. Their main result is a characterization of an optimal explore and exploit scheme.
In this work  we extend \cite{Kremer2014} by adding a layer of a social network according to which agents can observe each other. It turns out that when observability is factored in  the scheme proposed by Kremer et. al. is no longer incentive compatible. In our main result we provide a tight bound on how many other agents can each agent observe and still have an incentive-compatible algorithm and asymptotically optimal outcome. More technically, for a setting with $N$ agents where the number of nodes with degree greater than $N^{\alpha}$ is bounded by $N^{\beta}$ and $2\alpha+\beta <1$ we construct incentive-compatible asymptotically optimal mechanism. The bound $2\alpha+\beta <1$ is shown to be tight.

\end{abstract}
\parskip\medskipamount
\section{Introduction}

In a variety of settings members of a society are faced with a set of possible actions which rewards are unknown. Each agent chooses her action and  social learning may entail that many will choose the optimal action. This may be the case for selecting among alternative routes for traveling from one location to another, choosing a holiday destination, choosing a service provider (e.g., an accountant or an ISP) and more. In such settings the a-priori optimal alternative may often be a-posteriori inferior but nevertheless, as no one would like to experiment with an a-priori inferior alternative, society might converge on the wrong action resulting in a market failure.

A similar dilemma is central to on-line Explore \& Exploit paradigm [E\&E], a rich research area in Machine learning [ML] \cite{CesaBianchi}. In that setting, a central planner wants to learn the optimal action as soon as possible. To do so he can try out various actions and based on the history of results decide on whether to continue with experimentation or to exploit his knowledge. In this literature the central planner has full control over the experiment design and the history of results. A natural question is how this E\&E paradigm works out when the experiments are actually controlled by self motivated agents and the central planner can only make recommendations for actions.

When modeling society of agents where actions are taken in a decentralized fashion one typically accounts for some of the key aspects of the society. Such key aspects include the incentive structure of the individuals, the communication structure among the agents and the private prior information agents may hold. These three  modeling ingredients are central to the literature on social learning, where agents have initial conflicting beliefs on the optimal action but learn from each other while taking actions simultaneously and repeatedly. This literature has its roots in Aumann's agreeing to disagree \cite{Aumann76} and has been later extended in a variety of papers (e.g.,\cite{GaleKariv,RosenbergSV09,MuellerFrank,MosselST}).
The literature on herding also studies how these three components interact. In the herding literature, similar to the E\&E setting above, agents act sequentially and each agent acts once. Initially, more emphasis has been given in the herding literature to the subjective information structure (e.g., \cite{Banerjee,SmithSorensen,Bikhchandani,Alon2012}).%
\footnote{The mentioned work employ a game-theoretic setup. When restricted to convention evolution in pure coordination games, other aspects such as the design of adaptation rules come into play \cite{STCONV,Kittock}.}
However, recently the importance of the social network structure has been acknowledged (see \cite{Acemoglu} for a discussion of the observability assumption). In the herding literature one may also observe market failure despite of the fact that the collective holds the information required for making the optimal choice.

The first paper to marry the social aspects with the challenge of E\&E, a new research domain for which we coin the term {\em `social explore and exploit'}, is Kremer et. al. \cite{Kremer2014}.  In that work the authors introduce a naive setting and study optimal explore and exploit schemes that account for agents' incentives.%
 \footnote{We use the notion of a `naive setting' for settings where the optimal non-social explore and exploit scheme is trivial - try all actions sequentially, each once, and settle on the optimal one thereafter.} The following example is useful to understand the model of social explore and exploit due to Kremer et. al.

\begin{example}\label{Ex1}
Assume there are two routes, denoted T(rain) and R(oad) from point A to point B. The latencies in both alternatives are known to change on a daily basis. On each given day the travel time on R is a constant that is sampled from a uniform distribution on the interval $[0,6]$ (with a mean of 3 hours) whereas T is uniformly distributed on the interval $[1,3]$ (with a mean of 2 hours).
A benevolent dictator would like to make sure most agents take the faster option on any given day. To do so he would dictate to the first arriving agent each day to use option R and to the second one to use T after which he would surely know which is the better option on that day. Hence, as of the third agent he would dictate the a-posteriori optimal action. On the other hand, without any central mechanism, no self-motivated agent will try alternative R and so with a high probability all agents will use the inferior option each day
(with probability $Prob(R<T) = \frac{1}{3}$ to be precise). In the 'social E\&E' setting we introduce a central planner (one can think of it as a recommendation engine), which observes the outcome of the agents and makes a non-enforceable recommendation to subsequent agents on which action to take. Kremer et. al. show that the introduction of such a central planner can lead to choosing the optimal action even when agents are self motivated.
\end{example}

To be more specific, Kremer et. al. identify an incentive compatible scheme with which a central planner can asymptotically steer the users towards taking the optimal action. This exciting news has been extended in \cite{Mansour2015} to several more elaborate bandit settings and to additional optimization criteria such as regret minimization. An implicit assumption in both papers is that agents cannot see each other's behavior. This assumption turns out to be critical. In fact, even with very little observability, for example when each agent just sees his predecessor, the schemes proposed in both works cease to be incentive compatible and lead to market failure. In this work we investigate the conditions on the social network which allow for asymptotically optimal outcomes. Thus, we extend \cite{Kremer2014} by adding the additional layer of a social network and show conditions under which the essence of their results, albeit with a different mechanism, can still be maintained even though agents may observe each other.

Needless to say, the ability to observe peers' actions is realistic and quite common in many applications, ranging from route recommendation to hotel recommendation and transportation recommendation, etc. In any of these recommendation systems some exploration may be needed, and no one wishes to be the one to explore so others can benefit, while some observability of some others' recommended actions does exist.

Technically, we extend the `social explore and exploit' setting and incorporate the visibility of actions of peers in a social network, creating a more complete initial theory of economic recommendation systems. We incorporate into the model of \cite{Kremer2014} a notion of visibility, captured by a visibility graph. In the visibility graph agents are nodes, and an edge $(a,b)$ implies that agents $a$ and $b$ can observe each other's action. Our main result is that for a setting with $N$ agents, when the number of nodes with degree greater than $N^{\alpha}$ is bounded by $N^{\beta}$, where $2\alpha+\beta <1$,  there exists an incentive-compatible algorithm  leading to asymptotically optimal outcome. In other words, there exists a recommendation mechanism which agents will gladly follow and which ensures that a vanishing proportion of the agents  take a sub-optimal action. We also show  that the result is tight, in the sense that there is a visibility graph, where $2\alpha + \beta =1$ (In particular the complete graph) and approximately optimal outcome can not be obtained by any incentive-compatible algorithm.

\section{Model}
We consider a setting where agents arrive sequentially and must choose an action from a finite set $A$. The reward from taking action $a \in A$ is given by a commonly known non-atomic random variable $V_a$ taking values in some interval $I=[L,R]$ and is the same across all agents. Agents would like to maximize their value. At stage $0$ neither the agents nor the social planner know the value of the random variables $\{V_a\}_{a\in A}$. When agent $n$ arrives the social planner sees the choices made by agents $1,\ldots,n-1$ and the corresponding rewards and chooses to communicate some message, $m \in M$ (a recommendation), to the agent. Based on this message, as well as other information available to the agent, she chooses some action.

This extra information available to each agent is the actions chosen by some of his predecessors, those that he can observe. Let $\cal N$ denote the set of $N$ agents and for each $n\in \cal N$ let $B(n) \subset N \setminus \{n\}$ denote the set of friends of $n$. The agent arriving at stage $t$ gets to see the action (but not the reward) chosen by the subset of his friends that preceded him. Coupled with the message received from the social planner he must choose his action.

Formally, The strategy of the social planner at time $n$ is a function $\tilde M^n: (A \times I)^{n-1}\to M$.%
\footnote{Restricting the planner to pure strategies is done for the sake of simplicity only. It
is easy to see that each of the arguments in the following sections holds true when the
planner is also allowed to use mixed strategies, and that the resulting optimal strategy of
the planner is pure.}
Note that once some agent chooses the action $a$ the realization of $V_a$ is known thereafter to the social planner. Let agent $n$ be the $t$-th agent arriving and let agents $\{1,2,\ldots,t-1\}$ be his predecessors, then the strategy of agent $n$ depends on the message communicated to him by the planner and the actions taken by the set of agents  $B^t(n) = B(n)\cap\{1,2,\ldots,t-1\}$. That is, $n$'s strategy is represented by some function $\sigma_n:A^{B^t(n)} \times M \to A$.%
\footnote{Our model assumes that agents know their position in the sequel. However the results reported here extend to the case where agents arrive randomly and do not know their position in the sequel.}

The goal of each agent is to maximize her expected reward while the goal of the planner is to maximize the expected average reward, or the expected proportion of agents who choose an ex-post optimal action. Note that the social planner would like to induce agents to experiment with previously untried actions. Ideally, the social planner would like each of the first $|A|$ agents to experiment with a different action. Once that happens she can ensure all subsequent agents take the optimal action and asymptotically maximize the average reward. On the other hand, agents are short-sighted and do not want to experiment with ex-ante inferior actions. This tension is at the core of the analysis we provide.

Hereinafter we assume the set of alternatives is binary, $|A|=\{a,b\}$. Let $\mu_a,\mu_b$ be the expectations of $V_a,V_b$ correspondingly and assume, without loss of generality, that $\mu_a > \mu_b$. In what follows, similar to Kremer et al \cite{Kremer2014}, we provide a direct mechanism (one for which $M=A$) which is incentive compatible (IC) and asymptotically optimal. That is, agents comply with the mechanism's recommendation in equilibrium and only a vanishing proportion of players take the inferior action, for large enough $N$.

\subsection{The No Visibility Case}

We begin with the exact setting of Kremer et al \cite{Kremer2014} and assume no visibility across agents (`blind' agents). Formally this is the setting where $B(n) = \emptyset \ \ \forall n$.
The mechanism we formulate is different than that introduced in \cite{Kremer2014}. Whereas we do not know how to adapt the Kremer et al mechanism to the case with visibility ours can be adapted, as we do in subsequent sections.

The underlying idea in the following construction is for the mechanism to recommend action $a$ to the first agent, who happily complies. Thereafter the mechanism knows the value of $V_a$. The mechanism commits up front to a finite set  of agents to recommend action $b$ only when $V_a$ falls into some pre-defined set. For this to be incentive compatible the expected value of $V_a$, conditional on that set (which is the expectation of $V_a$ from the perspective of the agent that is recommended $b$) must not be greater than $\mu_b$. If the mechanism can ensure that the aforementioned sets cover all the possible realizations of $V_a$ then surely one agent at least will try action $b$. Note that these sets need not be disjoint and so it might be the case the more than one agent is recommended $b$.

These sets will be induced by some partition of the interval $I$ which is defined in the next Lemma:

\begin{lemma}\label{LEM1}
There exists a finite partition $\{D_k\}_{k=0}^K$ of $[L,R]$ such that
$D_0=(L,\mu_b)$, $E(V_a|D_0 \cup D_k) = \mu_b \ \forall 1\leq k \leq K-1$ and $E(V_a|D_0 \cup D_K) \leq \mu_b$. Furthermore,
$\frac { \mu_{a} - \mu_{b}}{p(V_{a} \in D_0)[\mu_{b} - E(V_{a}|D_0)]} + 1 \leq K \leq \frac { \mu_{a} - \frac{\mu_{b}}{2}}{p(V_{a} \in D_0)[\mu_{b} - E(V_{a}|D_0)]} + 2$
\end{lemma}

\textbf{Proof}
Consider the function $f_1(x)=E(V_a|D_0 \cup [\mu_b,\mu_b+x])$ defined for any nonnegative $x$.
Note that $f_1(0)< \mu_b$ and $f(R-\mu_b) = \mu_a > \mu_b$. As $f_1$ is continuous there exists, by the intermediate value theorem, a value $x_1$ for which $f_1(x_1)=\mu_b$. Set $D_1=[\mu_b,x_1)$ (in fact, $x_1$ is unique as $f_1$ is monotonic.

If $E(V_a | D_0 \cup [\mu_b,R] \setminus D_1) < \mu_b$ then set $K=2$ and
$D_2=[\mu_b,R] \setminus D_1$. Otherwise, consider the function
$f_2(x)=E(V_a|D_0 \setminus D_1 \cup [x_1,x_1+x])$
applying the intermediate value theorem as before, there must be some $x_2$ such that
$f_2(x_2)=\mu_b$. Set $D_2=[x_1,x_2)$.

Repeat this iteratively:
Assume $D_j=[x_{j-1},x_j)$ have been defined for $j=1,\dots,k-1$.
If $E(V_a | D_0 \cup [x_{k-1},R]) < \mu_b$ then set $K=k$, $D_K= [x_{k-1},R]$, and halt.
Otherwise, let $f_k(x) = E(V_a|D_0 \setminus \cup_{j=1}^{k-1} D_j \cup [x_{k-1},x_{k-1}+x])$.
By applying the intermediate value theorem as before, there must be some $x_k$ such that
$f_k(x_k)=\mu_b$. Set $D_k=[x_{k-1},x_k)$.

We now turn to argue that the above procedure eventually halts. To see this note that
$$
\mu_{b} = E(V_a|D_0 \cup D_k) =
\frac{P(V_{a}\in D_0)E(V_a|D_0)+ P(V_{a}\in D_k )E(V_a|D_k)}{P(V_{a}\in D_0)+P(V_{a}\in D_k )} \Rightarrow
$$
$$
P(V_{a}\in D_0)E(V_a|D_0)+ P(V_{a}\in D_k )E(V_a|D_k) = \mu_{b}[P(V_{a}\in D_0)+P(V_{a}\in D_k )] \Rightarrow
$$
$$
P(V_{a}\in D_k )[E(V_a|D_k)-\mu_{b}] = P(V_{a}\in D_0)[\mu_{b}-E(V_a|D_0)] \Rightarrow
$$
$$
P(V_{a}\in D_k )E(V_a|D_k) \geq P(V_{a}\in D_0)[\mu_{b}-E(V_a|D_0)].
$$
Note that the right hand side of the last inequality is some positive number, $\delta$, independent of $k$ and also that $E(V_a|D_k) \le R$. Hence $P(V_{a}\in D_k )\geq \frac{\delta}{R}.$

Summing over $k$: $1 \ge \sum_{k=0}^K P(V_{a}\in D_k ) \geq \sum_{k=0}^K \frac{\delta}{R}$ which implies that $K \leq \frac{R}{\delta}$

Let us now compute the upper bound on the value of $K$.

Note that since $E(V_a|D_0 \cup D_{K-1}) \geq E(V_a|D_0 \cup D_{K})\bigwedge E(V_a|D_K) \geq E(V_a|D_{K-1})$ we can conclude that $p(V_a \in D_K) \leq p(V_a \in D_{K-1})$. As $p(V_a \in D_K) + p(V_a \in D_{K-1}) \leq 1$ we get $p(V_a \in D_K) \leq \frac{1}{2}$. Therefore,
from the above we can conclude that:
\begin{equation}\label{eq1}
\Sigma_{k=1}^{K-1}P(V_a\in D_0 \cup D_{k}) E(V_{a}|D_0 \cup D_{k})=
\mu_b [(K-1) P(V_{a}\in D_0) + P(V_{a}\in D_1 \cup ...\cup D_{k-1})] =
$$
$$
\mu_b [(K-2) P(V_{a}\in D_0) + 1 - P(V_{a} \in D_K) \geq \mu_b [(K-2) P(V_{a}\in D_0) + \frac{1}{2}].
\end{equation}

On the other hand we may substitute
$E(V_{a}|D_0 \cup D_{k})$ with
$ \frac{ p(V_a \in D_0)E(V_a|D_0) +  p(V_a \in D_{k})E(V_a|D_{k})}{p(V_a\in D_0 \cup D_{k})}$, and so:

\begin{equation}\label{eq2}
\Sigma_{k=1}^{K-1} P(V_{a}\in D_0 \cup D_{k}) E(V_{a}|D_0 \cup D_{k})=
$$
$$
\Sigma_{k=1}^{K-1}   p(V_{a} \in D_0)E(V_{a}|D_0) +  p(V_{a} \in D_{k})E(V_{a}|D_{k})=
$$
$$
(K-1) p(V_{a} \in D_0)E(V_{a}|D_0) +\Sigma_{k=1}^{K-1} p(V_{a} \in  D_{k})E(V_{a}|D_{k}) =
$$
$$
(K-2)P(V_{a} \in D_0)E(V_{a}|D_0) + \mu_{a} -p(V_a \in D_K)E(V_{a} |D_K) \leq
$$
$$
(K-2)P(V_{a} \in D_0)E(V_{a}|D_0) + \mu_{a}
\end{equation}

From equations  \ref{eq1} and \ref{eq2}  we get:
$$
(K-2)P(V_{a} \in D_0)E(V_{a}|D_0) + \mu_{a} \geq  [(K-2)P(V_{a} \in D_0) + \frac{1}{2}] \mu_{b} \Rightarrow
$$
$$
K \leq \frac { \mu_{a} - \frac{\mu_{b}}{2}}{p(V_{a} \in D_0)[\mu_{b} - E(V_{a}|D_0)]} + 2.
$$

Finally, we also compute a lower bound on the value of $K$:%
 \footnote{Note that we make use for this lower bound when we study high visibility graphs in section \ref{section_high_visibility}.}
\begin{equation}\label{eq3}
\Sigma_{k=1}^{K}P(V_a\in D_0 \cup D_{k}) E(V_{a}|D_0 \cup D_{k}) \leq
\mu_b [(K) P(V_{a}\in D_0) + P(V_{a}\in D_1 \cup ...\cup D_{K})] =
$$
$$
\mu_b [(K-1) P(V_{a}\in D_0) + 1]
\end{equation}

On the other hand:

\begin{equation}\label{eq4}
\Sigma_{k=1}^{K} P(V_{a}\in D_0 \cup D_{k}) E(V_{a}|D_0 \cup D_{k})=
\Sigma_{k=1}^{K}   p(V_{a} \in D_0)E(V_{a}|D_0) +  p(V_{a} \in D_{k})E(V_{a}|D_{k})=
\end{equation}
$$
= K p(V_{a} \in D_0)E(V_{a}|D_0) +\Sigma_{k=1}^{K} p(V_{a} \in  D_{k})E(V_{a}|D_{k}) =
(K-1)P(V_{a} \in D_0)E(V_{a}|D_0) + \mu_{a}
$$
Combining equations  \ref{eq3} and \ref{eq4}  we get:
$$
(K-1)P(V_{a} \in D_0)E(V_{a}|D_0) + \mu_{a} \leq  [(K-1)P(V_{a} \in D_0) + 1] \mu_{b} \Rightarrow
$$
$$
K \geq \frac { \mu_{a} - \mu_{b}}{p(V_{a} \in D_0)[\mu_{b} - E(V_{a}|D_0)]} + 1.
$$
Q.E.D

A {\em direct revelation mechanism} is a mechanism for which the message space equals the action space, $M=\{a,b\}$. Given the partition $\{D_k\}_{k=0}^{K+1}$ we define the following direct revelation mechanism for our social planner:

\begin{framed}
\noindent{\bf No Visibility Mechanism}:
\begin{itemize}
\item
If $P(V_a< \mu_b) = 0$ then set $\tilde M^{n} = a$ $ \forall n$.
\item
If $ P(V_a< \mu_b) > 0$ then:
\begin{enumerate}
  \item set $\tilde M^{1} = a$
\item
For $n=2 \ldots,K+1$ let $\tilde M^{n}=b$ whenever
$V_a \in (D_0 \cup D_{n-1})$ and $\tilde M^{n}=a$ otherwise.
\item
Let $c\in A$ be the best action among those chosen by agents $1,\ldots, K+1$. For any agent $n>K+1$ set $\tilde M^n =  c$.
\end{enumerate}
\end{itemize}
\end{framed}

We now turn to argue that the {\it No Visibility Mechanism} is incentive compatible, that is each agent will use the action that is recommended to him by the planner. Hence one of the agents $2,\ldots,K+1$ will surely try action $b$. This, in turn, implies that all agents $n>K+1$ will be recommended the optimal action.

\begin{theorem}\label{THM1}
The {\it No Visibility Mechanism} is incentive compatible
\end{theorem}

\textbf{Proof:}
Since $\mu_a > \mu_b$ the first agent will clearly comply with the social planner's recommendation to take action $a$.

For each agent $2 \leq j \leq k+1$:
\begin{itemize}
 \item
 Note that the event $\tilde M^{j}=b$ is the same as the event $V_{a}\in (D_0 \cup D_{j-1})$. Thus, the expected reward from taking action $a$ is $E(V_{a}|(D_0 \cup D_{j-1})= \mu_b$, which is exactly the expected reward from taking action $b$. Agent $j$ will therefore be indifferent between $a$ and $b$ and might as well take action $b$ as the planner recommended.
 \item
The event $ \tilde M^{n}=a$  is equal the event $V_{a}\notin (D_0 \cup D_{n-1})$. However as $E(V_{a}|(D_0 \cup D_{n-1})= \mu_b <\mu_a$ we conclude that the expected reward from taking action $a$, given the message $\tilde M^{n}=a$ is $E(V_{a}|V_{a} \notin (D \cup D_{n-1})) > \mu_{b}$, where the expected reward from taking action $b$ is $\mu_b$. Therefore agent $j$ will prefer action $a$, as recommended by the planner.
\end{itemize}

Recall that $\cup_{k=0}^K D_k = [L,R]$ which implies that at least one agent will be recommended, and consequently choose, action $b$. Therefore, for any agent $j>K+1$ the planner recommends the optimal action and so agents will comply.

Q.E.D
\begin{definition}
Let $U_j$ be the utility of agent j. An incentive compatible direct revelation algorithm is {\em asymptotically optimal} if $ \forall \epsilon >0 , \exists \bar N$ such that $\forall N \geq \bar N, V_a, V_b \in I:\  \frac{\Sigma_{k=1}^{N}U_k}{N \max(V_a,V_b)} > 1- \epsilon$.
In words, the average utility of the agents goes to $max(V_a,V_b)$.
\end{definition}

 \begin{corollary}
   If $P(V_a< \mu_b) > 0$ then the {\it No Visibility Mechanism} is asymptotically optimal
  \end{corollary}

\textbf{Proof:} By Theorem \ref{THM1} the {\it No Visibility Mechanism} is incentive compatible and so after the first $K+1$ agents the social planner knows the values of both $V_a$ and $V_b$. This ensures that agents $k+2 ....N$ will take the optimal action. As $K$ is independent of the total number of agents, $N$, the proportion of agents taking the optimal action increases to one as $N$ grows.
\newline
   Q.E.D

   Note that this gives an alternative technique to \cite{Kremer2014}, which can be later generalized to address the case of network that allows for visibility.

\subsection {The Medium Visibility Case}

We next turn to study the case where all agents do have some visibility, albeit limited visibility.
In particular we assume that $|B(n)|\le N^{\alpha}\ \ \forall n$ and for some $\alpha <0.5$.
Unfortunately we cannot use the  {\it No Visibility Mechanism} as it may become non incentive compatible whenever $B(n) \not =\emptyset$ (at least for the first $K$ agents).  We turn to explain the underlying reasons we lose the IC(incentive compatible) property:
\begin{enumerate}
\item What happens when $k>j$, both are in $K$ and $j\in B(k)$?
 Consider an instance where $V_a \in D_k$. In that case
assuming $IC$, $j$ will take action $a$ and $k$ will be recommended action $b$.
From these two, agent $k$ concludes that $V_a \in (D_0 \cup D_k) \setminus (D_0\cup D_j) = D_k$, in which case he will take action $a$, contradicting IC.
\item What happens when $k>v>j>i$, where $i,j,k$ are in K but v is not in K, and both $i,j$ are in $B(v)$ while $v$ is in $B(k)$? Consider an instance where $V_a \in D_k$. In that case
assuming $IC$, $i$ and $j$ will take action $a$, $v$ will see that both $i$ and $j$ took action $a$ and will take action $a$ as recommended, and $k$ will be recommended action $b$. But $k$ can see that $v$ took action $a$. However if  $V_a \in D_0$ then assuming $IC$ both $j$ and $k$ will take action $b$ and since $v$ can see both of them he can conclude $V_a \in D_0$ and "herd" $b$. Therefore $k$ can conclude $V_a \in D_0 \cup D_k \setminus D_0 = D_k$ in which case he will take action $a$, contradicting IC.

\end{enumerate}

However, a variant of the {\it No Visibility Mechanism}, which we term the {\it Medium Visibility Mechanism}, works. The way we adapt to the medium visibility case is by choosing the set of K agents in a way that they do not see each other, directly or indirectly, which is why the original mechanism fails. This will entail an increase in the number of initial agents which are not necessarily recommended the optimal action from $K+1$ to a larger number, but nevertheless the asymptotic efficiency will still prevail.

To introduce this variant we use the following notation:   For a subset of agents $\tilde N \subset N$, $B(\tilde N) = \cup_{n\in \tilde N}B(n)$.
In words, $B(\tilde N)$ is the set of neighbors of  $\tilde N$.

The main idea behind the following algorithm is to find $K$ agents that cannot see each other, moreover that there is no possibility that any other agent (outside of those $K$ agents) will  be able to see two or more agents from this group, so no other agent can reflect the group choices to other agents from the group. Note that this algorithm is dynamic and we do not need to know the order of arrival in advance.

\begin{framed}
\noindent{\bf Medium Visibility Mechanism}:%
\begin{itemize}
\item
Let $M=\{a,b\}$
\item
If $P(V_a< \mu_b) = 0$ then $\tilde M^{n} = a \ \forall n$
\item
If $ P(V_a< \mu_b) > 0$ then set $\tilde M^{1} = a$, $\tilde\rho = \emptyset$, $k=0$, and $\tilde N = \emptyset$
\item For $n=2,\ldots,N$:
\begin{itemize}
  \item While $k<K$ do:
  \begin{itemize}
  \item If $n \in \{B(\tilde \rho) \cup B(B(\tilde \rho))\}$ then $\tilde M^{n} = a$ and $ \tilde N = \tilde N \cup \{n\}$.
  \item If $n \not \in \{B(\tilde \rho) \cup B(B(\tilde \rho))\}$ then $\tilde\rho=\tilde\rho\cup\{n\}$, $k=k+1$ and
      \begin{itemize}
      \item $\tilde M^{n}=b$ whenever $V_a \in (D_0 \cup D_{k})$
      \item $\tilde M^{n}=a$ otherwise.
      \end{itemize}
  \end{itemize}
  \item If $k=K$ then $\tilde M_{n}=\argmax_{a,b}(V_a,V_b)$.
\end{itemize}
\end{itemize}
\end{framed}

Note that the above mechanism essentially applies the {\it No Visibility Mechanism} to the subset $\tilde\rho$ of agents. In the process it `ignores' another set of agents, those that have high visibility, and are denoted $\tilde N$. The next lemma argues that the agents in $\tilde N$ have limited visibility into $\tilde\rho$:

\begin{lemma}
\label{Lemma_no_two_agents}
Any agent $n\in \tilde N$ sees at most one agent in $\tilde \rho$.
\end{lemma}

\textbf{Proof:}
Let us assume this is not true and that in fact there exist $i,j \in\tilde{\rho}$ such that $i<j$ and
 $i,j \in B(n)$. Let $i,j$ be the first two such agents to satisfy these requirements.
Note that  $n \in B(i)$ and $j\in B(n)$ which implies that $j \in B(B(i))$. This, in turn, is a contradiction to the fact that $j \in \tilde{\rho}$.

QED

\begin{theorem}\label{THM2}
The {\it Medium Visibility Mechanism} is incentive compatible.
\end{theorem}
\textbf{Proof:}
Similar to the {\it No Visibility Mechanism}, the first agent will get the message $a$ and will optimally comply.
We now consider 3 cases: an agent in $\tilde N$, an agent in $\tilde \rho$ and agents $j$ arriving when $k=K$.

\begin{enumerate}
\item
Consider an agent $n\in \tilde N$ who is necessarily recommended action $a$. Assume all other agents follow the recommendation of the mechanism.
By Lemma \ref{Lemma_no_two_agents}, $|B(n) \cap \Tilde \rho | \le 1$. Assume $|B(n) \cap \Tilde \rho | =0$ then agent $n$ has no other information above and beyond his prior and hence chooses action $a$ as he is recommended. If $|B(n) \cap \Tilde \rho | =1$ then the agent in $\tilde \rho$ that $n$ observes, say agent $j$, may have either taken action $a$ or $b$. In the former case $n$ infers that
$V_{a} \not \in (D_{0} \cup D_{j})$ which implies that $V_a$ is better than $b$ and so action $a$ is chosen.
In the latter case $n$  infers that  $V_{a} \in (D_{0} \cup D_{j})$, from which he can only conclude that the expected reward in both actions is equal and hence will also take action $a$.
\item
Consider an agent $j\in \tilde \rho$ and assume all other agents follow the recommendation of the mechanism.
 According to the {\it Medium Visibility Mechanism}, $j \notin B(\tilde \rho \setminus \{j\})$.  Therefore, all the predecessors observed by agent $j$ have received no information and so provide $j$ with no information themselves. Therefore, his expected reward from both actions, given the recommendation of the {\it Medium Visibility Mechanism} is the same as that of agent $j+1$ in the case $B(n) = \emptyset$ and a recommendation of the {\it No Visibility Mechanism}. Incentive compatibility of $j$ follows now from Theorem \ref{THM1} .
 \item
 Consider an agent $j$ arriving when $k=K$ and assume all other agents follow the recommendation of the mechanism.
  Recall that $\cup_{k=0}^K D_k = [L,R]$ which implies that at least one agent from $\tilde \rho$ chose action $b$. Therefore, the planner recommends agent $j$ the optimal action and so he will comply.
\end{enumerate}

Q.E.D \newline
\begin{theorem}\label{THM3}
If $|B(n)|\le N^{\alpha}\ \ \forall n$ then the value of the parameter $k$ of the {\it Medium Visibility Mechanism} terminates in $K$ whenever $N > 2(K-1)N^{2\alpha}$. Furthermore, in that case $|\tilde \rho \cup \tilde N | \le 2(K-1)N^{2\alpha}$.
\end{theorem}
\textbf{Proof:}
Assume the algorithm terminates with a value $j<K$. By the construction $|B(\tilde \rho)| \leq N^{\alpha}|\tilde \rho|$ and so $|B(B(\tilde \rho))| \leq N^{2\alpha}|\tilde \rho|$ at the termination. As $|\tilde \rho| = j$ there are at most $j(N^{\alpha}+N^{2\alpha})$ agents in $B(\tilde \rho)\cup B(B(\tilde \rho))$.
And so if there are more than $j(N^{\alpha}+N^{2\alpha})$ additional agents one must satisfy the conditions required to join $\tilde \rho$. However, this must hold true for any $j=1,\ldots,K-1$ whenever there are initially $j+\sum_{j=1}^{K-1} j(N^{\alpha}+N^{2\alpha}) \le 2(K-1)N^{2\alpha} $ agents. Hence a contradiction.

Q.E.D

\begin{corollary}
if $|B(n)|\le N^{\alpha}\ \ \forall n$  and $\alpha < 0.5$ then the {\it Medium Visibility Mechanism} is asymptotically optimal
\end{corollary}

\textbf{Proof:} By Theorems \ref{THM2} and \ref{THM3} the {\it Medium Visibility Mechanism} is incentive compatible for large enough $N$ and so after $k=K$  the social planner knows the values of $V_a$ and $V_b$. By Theorem \ref{THM3}  $k=K$ after at most $2(K-1)N^{2\alpha}$ agents. and so at most $2(K-1)N^{2\alpha}$ will take the inferior action. As $K$ is independent of the total number of agents, $N$, and $2\alpha <1$ the proportion of agents taking the optimal action increases to one as $N$ grows.
\newline
   Q.E.D

\subsection {The High Visibility Case}\label{section_high_visibility}

We next extend our results and mechanisms to the case where a limited number of agents may exhibit a high number of neighbors. By this we mean that there exist  agents for which $|B(n)|>N^{\alpha}$, however there are less than $N^{\beta}$ such agents, where $\alpha$ and $\beta$ are  non-negative parameters satisfying $2\alpha+\beta <1$. Note the the Medium Visibility case satisfies this restrictions as $\alpha<0.5$ and $\beta=0$ in the environment.

The {\it Medium Visibility Mechanism} offers a solution when $|B(n)|\le N^{\alpha}\ \ \forall n$. However it may fail whenever there is even a single agent $j$ where $|B(j)| > N^{\alpha}$. The failure is due to the fact that the algorithm may terminate while $k=1$, in which case the conditions of Theorem \ref{THM3} are not satisfied. As an example of such an outcome consider a star shaped graph and an arrival order where the central agent arrives last.%
\footnote{Recall the mechanism is not forward looking and hence does not know that the central agent comes last.}

However, a variant of the {\it No Visibility Mechanism} and the {\it Medium Visibility Mechanism} works. The way we adapt the mechanism is by replicating the set of $K$ sets many times. Recall that each of the original $K$ sets was a union of two sets - $D_0$ from the left hand side of the mean $\mu_b$ and some $D_k$ from the right hand side of $\mu_b$. This construction implies that whenever an agent is recommended $b$ but happens to see that some agent before him took action $a$ then he can conclude that $V_a>\mu_b$ and refuse to accept the recommendation. To remedy this we construct the replicas in such a way that there is no overlap of the left hand side of one set from one replica with the left hand side of another set from another replica. Thus, if a low visibility agent that was recommended $b$ by the mechanism sees some high visibility agent that has taken action $a$ he will not be `polluted' by their action and will comply with the recommendation to take action $b$. To achieve this the mechanism uses a given replica of $K$ sets as long as no high visibility agent arrives. When a high visibility agent arrives the mechanism moves to the next replica of $K$ sets.

Let us now turn to the construction.
Let $\{D^0_{0},\ldots,D_{0}^{N^{\beta}}\}$  be a partition of  $D_0 =[L,\mu_b)$ such that
$E(V_{a}|V_a \in D_{0}^{j})  =E(V_{a}|V_a \in D_0)$ and $p(V_{a} \in D_{0}^j) =  \frac{1}{N^{\beta}+1}P(V_{a}\in D_0)$ $\ \forall j=0,\dots,N^{\beta}$.%
\footnote {This is feasible as $V_a$ is non-atomic.}

Let $\{D_{1},D_{2},\ldots,D_{K'}\}$ be a partition of the segment $[\mu_{b},R]$ such that
$E(V_{a}|V_a \in D^j_{0} \cup D_{i}) = \mu_{b}$ $\ \forall j=0,\ldots,N^{\beta}, i=1,\ldots,K'$.
\begin{lemma}\label{LEM3}
$K' \leq \frac { \mu_{a} - \frac{\mu_{b}}{2}}{\mu_a - \mu_b}(N^{\beta}+1)(k-1) + 2$ where K is the number of segments from Lemma \ref{LEM1}.
\end{lemma}

\textbf{Proof:} The upper bound computed in Lemma \ref{LEM1} for $K$ now applies to $K'$. Hence, for any set $D_0^j$:
$$
K'  \leq \frac { \mu_{a} - \frac{\mu_{b}}{2}}{p(V_{a} \in D^j_{0})[\mu_{b} - E(V_{a}|V_{a} \in D^j_{0})]} + 2 =
\frac { \mu_{a} - \frac{\mu_{b}}{2}}{\frac{1}{N^{\beta}+1}p(V_{a} \in D_{0})[\mu_{b} - E(V_{a}|V_{a} \in D_{0})]} + 2 =
$$
$$
=\frac { \mu_{a} - \frac{\mu_{b}}{2}}{\mu_a - \mu_b}(N^{\beta}+1)\frac { \mu_{a} - \mu_{b}}{p(V_{a} \in D_{0})[\mu_{b} - E(V_{a}|V_{a} \in D_{0})]} + 2  \leq \frac { \mu_{a} - \frac{\mu_{b}}{2}}{\mu_a - \mu_b}(N^{\beta}+1)(K-1) + 2,
$$
where the last inequality follows from the lower bound on $K$ computed in Lemma  \ref{LEM1}.

Q.E.D

Fix some parameter $\alpha$ and let $T = \{n : |B(n)|\le N^{\alpha}\}$.
 Let $S$ be the remaining set of agents and assume $\beta$ satisfies $|S| \leq N^{\beta}$.

The following mechanism is a variant of the {\it Medium Visibility Mechanism}. As usual, the first agent takes action $a$ and $V_a$ is revealed. At some point an agent is chosen as a candidate for a dynamic message (all others get the action $a$). This agent should not be a (first or second order) neighbor of any previous such agent. In contrast with the {\it Medium Visibility Mechanism}, the notion of neighbor we use is a neighbor in the sub-graph induced by the set $T$.
The message to this agent depends on whether or not $V_a \in D_0^z \cup D_i$, where $i$ is the counter of the candidates and $z$ is the counter for the number of agents from $S$ that have appeared so far. The extra trick we use here is that whenever an agent with many neighbors arrives we use a new sub-segment of $D_0$ and as a result candidate agents cannot conclude anything from observing agents is $S$.

Let us denote by $B_T(j) = B(j) \cap T$, the neighbors of $j$ in $T$, and naturally extend this to sets as follows, $B_T(\bar N) = \cup_{n \in \bar N}B_T(n)$.

\begin{framed}
\noindent{\bf High Visibility Mechanism:}%
Let $M=\{a,b\}\times\{TRUE, FALSE, SPECIAL\}$. If $P(V_a< \mu_b) = 0$ then $\tilde M^{n} = (a, TRUE) \ \forall n$. Otherwise:
\begin{enumerate}
  \item set $\tilde M^{1} = (a,TRUE)$
  \item set $experiment = TRUE$, $z=0$, $\tilde \rho= \emptyset$, $k=0$, $knowledge = FALSE$ and $c=a$
  \item For agents $n=2,\ldots,N$:
  \begin{itemize}

   \item if ($experiment = TRUE$)

    While $k<K'$ do:
    \begin{itemize}
        \item If $n \in S$ then
        \begin{itemize}
           \item let $z=z+1$
           \item if $knowledge = TRUE$:
            \begin{itemize}
               \item set $experiment = FALSE$.
               \item set $c=\argmax_{a,b}(V_a,V_b)$.
               \item $\tilde M_{n}=(c,FALSE)$
            \end{itemize}
               \item if $knowledge = FALSE$:
            \begin{itemize}
               \item $\tilde M_{n}=(a, TRUE)$.
            \end{itemize}
        \end{itemize}
        \item If $n \in T$ then
        \begin{itemize}
            \item If $n \notin B_T(\tilde \rho) \cup B_T(B_T(\tilde \rho))$ then $\tilde \rho =\tilde \rho \cup \{n\}$ and $k=k+1$.
            \begin{itemize}
                \item If $V_a \in D_0^z \cup D_k$ then $\tilde M_{n}=(b,TRUE)$ and set $knowledge = TRUE$.
                \item Otherwise  $\tilde M_{n}=(a,TRUE)$
            \end{itemize}
            \item If $n  \in B_T(\tilde \rho) \cup B_T(B_T(\tilde \rho))$ then $\tilde M_{n}=(a,TRUE)$.
        \end{itemize}
    \end{itemize}
      If $k =K'$ then:
    \begin{itemize}
       \item if $knowledge = TRUE$:
            \begin{itemize}
               \item set $experiment = FALSE$.
               \item set $c=\argmax_{a,b}(V_a,V_b)$
               \item $\tilde M_{n}=(c,FALSE)$
            \end{itemize}
       \item if $knowledge = FALSE$: (special case where $V_a \in D_0^z \cup...\cup D_0^{N^{\beta}})$
           \begin{itemize}
               \item $\tilde M_{n}=(b,SPECIAL)$
               \item $knowledge = TRUE$
            \end{itemize}
    \end{itemize}
    \item if ($experiment = FALSE$) then $\tilde M_{n}=(c,FALSE)$
  \end{itemize}
\end{enumerate}
\end{framed}

\begin{theorem}\label{THM4}
The {\it High Visibility Mechanism} is incentive compatible.%
\footnote{As the message space is formally not equal the action space what we clearly mean by IC is that agents will comply with the first component of the emessage, which is an action.}

\end{theorem}
\textbf{Proof:}
Similar to the {\it No Visibility Mechanism} and the {\it Medium Visibility Mechanism}, the first agent will get the message $(a,FALSE)$ and will optimally comply.
We now consider 7 cases:
\begin{enumerate}
\item{\bf Case 1: (experiment = TRUE, $k < K'$, $n \in S$ and $knowledge = FALSE$ )}: In this case the mechanism recommended  $(a,TRUE)$. Assume all other agents follow the recommendation of the mechanism. Since $V_b$ is not known to the planner it is obvious that all the actions that agent $n$ can see are $a$. And since $\mu_a > \mu_b$ agent $n$ will take action $a$ as recommended.
\item {\bf Case 2: (experiment = TRUE, $k < K'$, $n \in S$ and $knowledge = TRUE$)}(the first arriving agent from S following knowledge = TRUE):Assume all other agents follow the recommendation of the mechanism. However in the case the {\it High Visibility Mechanism} will set $experiment=FALSE$ and compute the optimal action before sending the message to $n$. The message sent to agent $n$ will contain $FALSE$ which will inform him that the experiment phase is over.  Whenever the flag in the message is FALSE (experiment phase over) the planner recommends agent n the optimal action and so he will comply.
\item {\bf Case 3: (experiment = TRUE, $k < K'$, $n \in T$ and $n\in \tilde \rho$)}. Assume all other agents follow the recommendation of the mechanism. Let us consider the following options:
\begin{itemize}
   \item Agent $n$ sees at least one agent from $S$ that took action $b$: Note that Assuming all other agents follow the recommendation of the mechanism according to the {\it High Visibility Mechanism} an agent from $S$ will take action $b$ only if $V_b = max(V_a,V_b)$ and the experiment flag will set to FALSE. In that case, however experiment = FALSE (see case 7 below)
   \item Agent $n$ does not see any agent from $S$ that took action $b$:
   \begin{itemize}
      \item $\tilde M_{n} = (a, TRUE):$ Given agent $n$ can not see any agent from $S$ that took action $b$, following the proof of theorem \ref{THM2} where $j \in \tilde \rho$ will prove incentive compatibility.
      \item $\tilde M_{n} = (b, TRUE):$ Whenever an agent from S arrives we use a new sub-segment of $D_0$ (change from $D_0^z$ to $D_0^{z+1})$ and as a result it is obvious that if $V_a\in D_0^z \cup D_i$ then he can not see any agent from $S$ that took action $b$ and therefore neither the agents from $S$ nor the TRUE flag add him any additional information. Therefore following the proof of theorem \ref{THM2} where $j \in \tilde \rho$ will prove incentive compatibility.
   \end{itemize}
\end{itemize}
\item {\bf Case 4: (experiment = TRUE, $k < K'$, $n \in T$ and $n\notin \tilde \rho$)}: Assume all other agents follow the recommendation of the mechanism. Let us consider the following options:
\begin{itemize}
    \item Agent $n$ sees at least one agent from $S$ that took action $b$: Note that assuming all other agents follow the recommendation of the mechanism according to the {\it High Visibility Mechanism} an agent from $S$ will take action $b$ only if $V_b = max(V_a,V_b)$ and the experiment flag will set to FALSE. In that case, however experiment = FALSE (see case 7 below)

    \item Agent $n$ does not see any agent from $S$ that took action $b$: In that case following the proof of theorem \ref{THM2} where $n \in \tilde N$ will prove incentive compatibility.
\end{itemize}
\item {\bf Case 5: (experiment = TRUE, $k = K'$, and $knowledge = FALSE$)}: In this case the mechanism recommended $(b,SPECIAL)$.  Assume all other agents follow the recommendation of the mechanism. Since $\cup_{i=1}^{K'}D_i = [\mu_b,R]$ We get that this special case is reached if and only if $V_a \in \cup_{z=z}^{N^{\beta}}D_0^z$. But $E(V_a|V_a \in \cup_{z=z}^{N^{\beta}}D_0^z) \leq \mu_b$. This implies that $b$ is at least as good as $V_a$. So agent that receives "special" in his recommendation will comply.
\item {\bf Case 6: (experiment = TRUE, $k = K'$, and $knowledge = TRUE$)}:Assume all other agents follow the recommendation of the mechanism. If $k=K'$ and $knowledge = TRUE$ the mechanism finish the experiment phase by setting $experiment = FALSE$ and the message sent to agent $n$ and to all the following agents will contain $FALSE$.  Whenever the flag in the message is FALSE (experiment phase over) the planner recommends agent n the optimal action and so he will comply.
\item {\bf Case 7: (experiment = FALSE)}: The message sent to the agent will contain $FALSE$. Assume all other agents follow the recommendation of the mechanism. Whenever the flag is FALSE (experiment phase over) the planner recommends agent n the optimal action and so he will comply.
\end{enumerate}

Q.E.D

\begin{theorem}\label{THM5}
Fix some parameter $\alpha$ and let $T = \{n : |B(n)|\le N^{\alpha}\}$.
 Let $S$ be the remaining set of agents and assume $\beta$ satisfies $|S| \leq N^{\beta}$. Then the {\it High Visibility Mechanism} will set $experiment = FALSE$ after at most $3K(\frac { \mu_{a} - \frac{\mu_{b}}{2}}{\mu_a - \mu_b}N^{\beta+2\alpha})$ agents, where K is the number from Lemma \ref{LEM1}.

\end{theorem}
\textbf{Proof:} \newline
We can deduce from the proof of Theorem \ref{THM3} that at most $2(K'-1)N^{2\alpha}$ agents from $T$ arrive before the parameter $k$ of the {\it High Visibility Mechanism} takes the value $K'$. Since $|S| \leq N^{\beta}$ there are at most $N^{\beta}$ agents from $S$ that arrive before $k=K'$.
Once $k=K'$ there could possibly be one extra agent until the parameter {\it experiment} takes on the value
{\it FALSE}. Therefore, the {\it High Visibility Mechanism} will set {\it experiment = FALSE} after less than $2(K'-1)N^{2\alpha} + N^{\beta} + 1$ agents. However, according to Lemma   \ref{LEM3}, $K' \leq \frac { \mu_{a} - \frac{\mu_{b}}{2}}{\mu_a - \mu_b}(N^{\beta}+1)(K-1) + 2$. Therefore the maximal number of agents that arrive before {\it experiment = FALSE} is $2(K'-1)N^{2\alpha} + N^{\beta} + 1 \leq 2(\frac { \mu_{a} - \frac{\mu_{b}}{2}}{\mu_a - \mu_b}(N^{\beta}+1)(K-1) + 2-1)N^{2\alpha} + N^{\beta} + 1 \leq 3K(\frac { \mu_{a} - \frac{\mu_{b}}{2}}{\mu_a - \mu_b}N^{\beta+2\alpha})$.

Q.E.D \newline
\begin{corollary}\label{CLR3}
Fix some parameter $\alpha$ and let $T = \{n : |B(n)|\le N^{\alpha}\}$.
Let $S$ be the remaining set of agents and assume $\beta$ satisfies $|S| \leq N^{\beta}$.  If $\beta + 2\alpha < 1$ then the {\it High Visibility Mechanism} is asymptotically optimal
\end{corollary}
\textbf{Proof:} \newline
As proved in theorem \ref{THM4} and \ref{THM5} the {\it High Visibility Mechanism} is incentive compatible and assure experiment = FALSE after at most $3K(\frac { \mu_{a} - \frac{\mu_{b}}{2}}{\mu_a - \mu_b}N^{\beta+2\alpha})$agents. And so at most $3K(\frac { \mu_{a} - \frac{\mu_{b}}{2}}{\mu_a - \mu_b}N^{\beta+2\alpha})$ will take the inferior action. As $K, \mu_{a},\mu_{b}$ are independent of the total number of agents, $N$, and since $\beta + 2\alpha < 1$ the proportion of agents taking the optimal action increases to one as $N$ grows.
   Q.E.D \newline

\subsection{The Very High Visibility Case}

As we have seen the {\it High Visibility Mechanism} works well with social networks where $T = \{n : |B(n)|\le N^{\alpha}\}$ and $S$,  the remaining set of agents, satisfies $|S| \leq N^{\beta}$, as long as $2\alpha + \beta < 1$. What happens if the social network fails to satisfy this visibility requirements. In our next Theorem we argue, via an example, that IC and efficiency cannot be guaranteed for such very high visibility networks.

In particular we demonstrate the impossibility for the case $\alpha=0$ and $\beta=1$, which implies that all agents can see each other:

\begin{theorem}\label{THM6}
Let $B(n)=N$ for all $n$ and assume $E(V_a|V_a<x)> \mu_{b}$, where $x=\inf\{y:Prob(V_b<y)=1\}$.
Then no incentive compatible asymptotically efficient mechanism exists.%
\footnote{Note that whenever $E(V_a|V_a<x)\le \mu_{b}$ a simple IC asymptotically efficient mechanism exists: After observing $V_a$ due to the first agent, whenever $V_a \geq  x $ recommend $a$ to all agents (this is efficient). Otherwise let the second agent know that $V_a<x$ in which case he is happy to experiment with $b$.}
\end{theorem}
\textbf{Proof:}
Assume an IC and asymptotically efficient mechanism exists. Let $W_{n} \subset I$ be the set for which agent $n$ is the first agent that is recommended to experiment with  $b$. As the mechanism is asymptotically efficient it must be the case that for any value if $V_a \le x$ some agent  will experiment with $b$, hence
$[L,x) \subset \cup_n W_n \subset I$, which implies
\begin{equation}\label{eq6}
E[V_a| V_a \in \cup_n W_n] \ge  E(V_a|V_a<x)> \mu_{b}.
\end{equation}
IC, coupled with the fact that agent $n$ can observe all past agents implies $E[V_a| V_a \in W_n \setminus \cup_{j=1}^{n-1} W_j] \le \mu_b$ for all $n$. Hence
$E[V_a| V_a \in \cup_n (W_n \setminus \cup_{j=1}^{n-1} W_j)] \le \mu_b$.  Note that
$\cup_n (W_n \setminus \cup_{j=1}^{n-1} W_j) = \cup_n W_n$ and so
$E[V_a| V_a \in \cup_n W_n] \le \mu_b$, contradicting inequality \ref{eq6}.

Q.E.D


\appendix

\end{document}